\documentstyle[twocolumn,aps,psfig]{revtex}
\begin{document}
\tighten
\draft
\newcommand{\ds}{\displaystyle}
\newcommand{\be}{\begin{equation}}
\newcommand{\en}{\end{equation}}
\newcommand{\bea}{\begin{eqnarray}}
\newcommand{\ena}{\end{eqnarray}}
\title{BTZ black hole from (3+1) gravity}
\author{Mauricio Cataldo $^{a}$
{\thanks{E-mail address: mcataldo@alihuen.ciencias.ubiobio.cl}},
Sergio del Campo $^{b}$  {\thanks{E-mail address:
sdelcamp@ucv.cl}} and Alberto Garc\'{\i}a $^{c}$ {\thanks{E-mail
address: aagarcia@fis.cinvestav.mx} }}
\address{$^a$
Departamento de F\'\i sica, Facultad de Ciencias, Universidad del
B\'\i o-B\'\i o, Avda. Collao 1202, Casilla 5-C, Concepci\'on,
Chile.
\\ $^b$ Instituto de F\'\i sica, Facultad de Ciencias
B\'asicas y matem\'aticas, Universidad Cat\'olica de Valpara\'\i
so, Avenida Brasil 2950, Valpara\'\i so, Chile.
\\
$^c$ Departamento de F\'\i sica, Centro de Investigaci\'on y de
Estudios Avanzados del IPN, Apart. Postal 14-740,
C.P.07000,M\'exico, D.F. M\'exico.}
\maketitle
\begin{abstract}
{\bf {Abstract:}} We propose an approach for constructing spatial
slices of (3+1) spacetimes with cosmological constant but without
a matter content, which yields (2+1) vacuum with $\Lambda$
solutions. The reduction mechanism from (3+1) to (2+1) gravity is
supported on a criterion in which the Weyl tensor components are
required to vanish together with a dimensional reduction via an
appropriate foliation. By using an adequate reduction mechanism
from the Pleba\'nski-Carter[A] solution in (3+1) gravity, the
(2+1) BTZ solution can be obtained.

PACS number(s): {04.20.Jb, 04.70.Bw }
\end{abstract}
\smallskip\
Lower dimensional gravity theory has attracted the attention of
the scientific community in the sense that one may use it as a
theoretical laboratory for studying certain aspects of gravity and
also because it may provide some clues for solving some open
problems in (3+1) dimensional gravity. On the other hand, it would
be of considerable interest to establish a link between the
physically relevant (3+1) gravity, via certain limiting process,
and the (2+1) gravity. This is the kind of problem, to which this
paper is addressed to.

In (3+1) dimensional gravity, black holes are the objects that have
received a lot of attention although their properties at the
quantum level remain until now as a mystery. It is believed that
black holes in (2+1) dimensions will provide a new insight toward a
better understanding of their properties in a four dimensional
spacetime.

It is well known that the Weyl tensor vanishes in (2+1) dimensions
and that the Riemann tensor can be written explicitly in terms of
the Ricci tensor and the scalar curvature $R$. Hence, vacuum
solutions to the Einstein equations are locally flat. When matter
sources and cosmological constant are present this is not longer
true~\cite{Frolov}. A criterion for obtaining (2+1) vacuum
solutions with cosmological constant from (3+1) solutions with
$\Lambda$ is based on the vanishing of the (3+1) Weyl tensor
together with a reduction of the number of dimensions using an
appropriate slicing at a constant spatial coordinate.

For instance, the reduction of the Pleba\'nski-Carter [A] metric,
after equating the Weyl tensor components to zero and accomplishing
the spatial slice at a constant value of a spatial coordinate,
yields a metric which can be thought of as the BTZ black hole
solution. The well known three dimensional spinning black hole was
found by Ba\~nados, Teitelboim and Zanelli (BTZ)\cite{BaTeZa}. The
BTZ solution with mass $M$, angular momentum $J$, and cosmological
constant $\Lambda=-1/l^2$ may be written in the form
\begin{eqnarray}
\label{metrica BTZ}
\ds ds^2= -N^2dt^2 + N^{-2}dr^2 + r^2 (N^{\phi} dt + d \phi)^2,
\end{eqnarray}
with the following metric functions:
\begin{eqnarray}
N^2= -M-\Lambda r^2+J^2/4r^2, \, N^{\phi}=-J/2r^2,
\end{eqnarray}
The function $N^{\Phi}$ is determined up to an additive constant,
say $A$, which can be eliminated by a linear transformation of the
$\phi$ coordinate.

The metric~(\ref{metrica BTZ}) remains as a solution disconnected
from the (3+1) dimensional world. Up to now, from the geometrical
and physical point of view, the BTZ black hole has no relation with
those geometries describing a black hole in four dimensions. A
question that one may pose is: could the BTZ black hole be obtained
from a ( 3+1) dimensional solution? The main objective of the
present paper is to give an answer to this question, which occurs
to be positive.

In order to establish this assertion, we consider the stationary
Einstein-Maxwell axisymmetric Pleba\'nski-Carter [A]
metric\cite{PlCa,GaMa}, which may be written as
\begin{eqnarray}
\label{Plebanski}
ds_{_{3+1}}^2= \frac{\Delta}{P} dp^2+ \frac{P}{\Delta}(d\tau+q^2
d\sigma)^2+
\frac{\Delta}{Q} dq^2  \nonumber \\
- \frac{Q}{\Delta} (d\tau-p^2 d \sigma)^2,
\end{eqnarray}
with structural functions
\begin{eqnarray}
\label{parametros}
P= \gamma - \overstar{g}^2 + 2 l p-
\epsilon p^2-\Lambda p^4,
\nonumber
\\ Q=\gamma + e^2 -2mq + \epsilon q^2 -\Lambda q^4 , \nonumber \\
\Delta= p^2 + q^2,
\label{Funs}
\end{eqnarray}
where the parameters $m$, $l$, $e$, $\overstar{g}$, and $\Lambda$
correspond to mass, magnetic mass, electric charge, magnetic
charge, and cosmological constant
($\Lambda=\lambda/3$),respectively,$\gamma$ is related with the
angular momentum, and $\epsilon$ is a parameter describing
geometrical properties.

The only nonvanishing Weyl component for the
metric~(\ref{Plebanski}) reads\cite{GaMa}
\begin{eqnarray}
\label{Weyl}
\ds \Psi_2=- \frac{(m+il)(q-ip)-e^2-\overstar{g}^2}{(q+ip)^3(q-ip)}.
\end{eqnarray}
The corresponding electromagnetic field for this metric, see
~\cite{GaMa}, can be written as
\begin{eqnarray}
E+i \overstar{B}=\frac{(p+iq)(e+i \overstar{g})}{\Delta^2}.
\end{eqnarray}

Our criterion requires that all components of the Weyl tensor
corresponding to a four dimensional solution must vanish. Since in
our case the number of the complex components of the Weyl tensor
reduces to one, by equating the expression~(\ref{Weyl}) to zero,
one has
\begin{eqnarray}
mq + lp + i(lq-mp) - e^2 - \overstar{g}^2=0,
\end{eqnarray}
therefore
\begin{eqnarray}
\label{condiciones para Weyl}
m=l=0,\, \,\hspace{1.0cm} e=\overstar{g}=0.
\end{eqnarray}
Notice that the electromagnetic field vanishes, hence the
corresponding conformally flat (3+1) solution occurs to be
uncharged. Assuming the fulfillment of the
conditions~(\ref{condiciones para Weyl}), the resulting structural
functions from~(\ref{parametros}), are
\begin{eqnarray}
\label{AA}
P= \gamma - \epsilon p^2-\Lambda p^4,
\nonumber \\
Q=\gamma + \epsilon q^2 -\Lambda q^4 , \\
\Delta= p^2 + q^2. \nonumber
\end{eqnarray}

For future convenience, we introduce the parameters
\begin{eqnarray}
\gamma = a^2,  \,\,\,\,\, \epsilon =-M,
\end{eqnarray}
which, as we shall see later, are related with the parameters
appearing in the BTZ metric. With these identifications the
resulting conformally flat (3+1) metric~(\ref{Plebanski}) becomes
\begin{eqnarray}
\label{3+1 BTZ}
\ds ds_{_{3+1}}^2= - \frac{\frac{a^2}{4q^2}-M- \Lambda 
q^2}{1+\frac{p^2}{q^2}} \left(d
\tau- 2 \frac{p^2}{a} d\phi \right)^2 \nonumber \\ + 
\frac{1+\frac{p^2}{q^2}}{\frac{a^2}{4
q^2} -M -\Lambda q^2} dq^2 +
\frac{p^2+q^2}{\frac{a^2}{4}+M p^2 -\Lambda p^4} dp^2
\\ \nonumber
\ds + q^2 \left ( \frac{1+ \frac{4 p^2}{a^2}(M-\Lambda p^2)}
{1+\frac{p^2}{q^2}} \right)
\left (\frac{a}{2 q^2} d \tau+ d \phi \right )^2,
\end{eqnarray}
where we have defined $\phi=a \sigma/2$. It is worthwhile to point
out that this metric~(\ref{3+1 BTZ}) occurs to be flat if the
cosmological constant $\Lambda$ vanishes. Since all components of
the Riemann tensor are zero, therefore via coordinate
transformations it can be reduced to the Minkowski spacetime.

The next step in our reduction process consists in requiring the
slicing  $p=constant$, ($p=const.$--foliation). For convenience, we
demand the following correspondence: $q \longrightarrow r$ and $d
\tau
\longrightarrow dt + 2 (p^2/a) d \phi$, therefore
from~(\ref{3+1 BTZ}) one gets
\begin{eqnarray}
\label{BTZ intermedio} \ds
ds_{_{2+1}}^2=\ds-\frac{\frac{a^2}{4r^2}-M-\Lambda r
^2}{1+\frac{p^2}{r^2}} dt^2 +
\frac{1+\frac{p^2}{r^2}}{\frac{a^2}{4 r^2}-M-\Lambda r^2} dr^2
\nonumber \\ \ds + r^2 \left( \frac{1+ \frac{4p^2}{a^2} (M
-\Lambda p^2)}{1 + \frac{p^2}{r^2}} \right) \left(\frac{a}{2 r^2}
dt+ \left( 1+\frac{p^2}{r^2} \right ) d \phi \right)^2.
\end{eqnarray}
For this foliation $p=const. \neq 0$, the metric~(\ref{BTZ
intermedio}) could be considered as a solution. We will postpone
the study of this interesting situation in a forthcoming
communication.

The most relevant case, which gives a positive answer to the posed
above question, is the slice $p=0$ of the (3+1) metric~(\ref{BTZ
intermedio}), together with the identifications $a \longrightarrow
-J$ and $\Lambda \longrightarrow
- 1/ l^2$, which yields
\begin{eqnarray}
\ds
ds_{_{2+1}}^2=- \left(\frac{J^2}{4r^2}-M+\frac{1}{l^2} r^2
\right)dt^2
\nonumber \\
+ \left(\frac{J^2}{4 r^2}-M+\frac{1}{l^2}r^2 \right)^{-1}dr^2 + r^2
\left(\frac{-J}{2 r^2} dt+  d \phi \right)^2.
\end{eqnarray}
This metric coincides just with the (2+1) BTZ black hole solution.
Summarizing, by applying our criterion (the vanishing of all
components of the Weyl tensor) and a slicing procedure ($p=0$--
foliation) to the (3+1) Pleb\'anski-Carter [A] solution, one is
able to arrive at the BTZ black hole in (2+1) dimensions.

Another point of view to the same problem consists in the search of
solutions to Einstein's field equations for a more general metric
than the metric~(\ref{Plebanski}). The line element to be
considered here is the stationary axisymmetric metric
\begin{eqnarray}
\label{Plebanski2}
ds_{_{3+1}}^2= \frac{\overline{\Delta}}{P(p)} dp^2+
\frac{P(p)}{\overline{\Delta}}(d\tau+N(q)
\, d\sigma)^2+ \frac{\overline{\Delta}}{Q(q)} dq^2  \nonumber \\
- \frac{Q(q)}{\overline{\Delta}} (d\tau+ M(p)\, d \sigma)^2,
\end{eqnarray}
where $P(p)$, $Q(q)$, $M(p)$, $N(q)$, and $\overline{\Delta}=
M(p)-N(q)$ are functions to be determined later on. For this
metric, using the Newman-Penrose tetrad formalism~\cite{KrStMcHe},
one defines the null tetrad
\begin{eqnarray}
e^{1}=\frac{1}{\sqrt{2}} \left[
\sqrt{\frac{\overline{\Delta}}{P}}dp + i
\sqrt{\frac{P}{\overline{\Delta}}} (d \tau + N d \sigma)   \right], \\
e^{2}=\frac{1}{\sqrt{2}} \left[
\sqrt{\frac{\overline{\Delta}}{P}}dp - i
\sqrt{\frac{P}{\overline{\Delta}}} (d \tau + N d \sigma)   \right], \\
e^{3}=\frac{1}{\sqrt{2}} \left[
\sqrt{\frac{\overline{\Delta}}{Q}}dq +
\sqrt{\frac{Q}{\overline{\Delta}}} (d \tau + M d \sigma)   \right], \\
e^{4}=\frac{1}{\sqrt{2}} \left[-
\sqrt{\frac{\overline{\Delta}}{Q}}dq +
\sqrt{\frac{Q}{\overline{\Delta}}} (d \tau + M d \sigma)   \right].
\end{eqnarray}
 From the second Cartan structure equations one arrives at the
expressions for traceless Ricci tensor $S_{a b}=
R_{ab}-\frac{1}{4}R g_{ab}$. By equating the components $S_{11}$,
$S_{33}$ and Im $S_{13}$ to zero, one arrives at a single equation
\begin{eqnarray}
\label{MN}
\ddot{M} + N^{''}=0,
\end{eqnarray}
on the other hand, $S_{12}=0$ leads to the equation
\begin{eqnarray}
\label{R01}
\ddot{P}-\frac{4 P}{\overline{\Delta}} \dot{P}+ 4
\frac{P}{\overline{\Delta}}- Q^{''} + \frac{4 q}{\overline{\Delta}} Q^{'} - 4
\frac{Q}{\overline{\Delta}}=0,
\end{eqnarray}
and, from the scalar curvature one obtains
\begin{eqnarray}
\label{escalar}
\ddot{P}+ Q^{''}+12 \Lambda \overline{\Delta}=0,
\end{eqnarray}
where dots and primes denote derivatives with respect to $p$ and
$q$, respectively. The nonvanishing complex Weyl coefficients are
\begin{eqnarray}
\Psi_{1}= - \Psi_{3}= i \frac{1}{4 \overline{\Delta}} \sqrt{PQ}
\left[\left( \frac{\dot{M}}{\overline{\Delta}} \right)^{.} -
\left(\frac{N^{'}}{\overline{\Delta}}  \right)^{'}\right]
\end{eqnarray}
and
\begin{eqnarray}
\label{Psi2}
\Psi_{2}= \frac{1}{12 \overline{\Delta}} \left\{\ddot{P}-3 
\frac{\dot{M}}{\overline{\Delta}}
\dot{P} - 2P \left[\frac{\ddot{M}}{\overline{\Delta}} -
2 \left(\frac{\dot{M}}{\overline{\Delta}}  \right)^2  \right.
\right.
\nonumber \\
\left. + \left(
\frac{N^{'}}{\overline{\Delta}} \right)^2 \right] +
Q^{''}+3 \frac{N^{'}}{\overline{\Delta}} Q^{'} \nonumber \\ \left.
+ 2Q
\left[  \frac{N^{''}}{\overline{\Delta}}+ 2\,
\left(\frac{N^{'}}{\overline{\Delta}}\right)^2-\left(\frac{\dot{M}}
{\overline{\Delta}}\right)^2 \right ] \right \} \nonumber \\ i
\left \{ \frac{N^{'}}{\overline{\Delta}}  \left (\dot{P}-2 \frac{\dot{M}}
  {\overline{\Delta}}P     \right) + \frac{\dot{M}}{\overline{\Delta}}
\left( Q^{'} + 2 \frac{N^{'}}{\overline{\Delta}} Q  \right)
\right \}.
\end{eqnarray}
To search the solution of this system of equations we shall first
determine $M(p)$ and $N(q)$. From equation~(\ref{MN}) one obtains
the polynomials $M(p)=\alpha p^2+\beta p +\gamma$ and $N(q)=
-\alpha q^2+\delta q +\rho$. The equation arising from the
vanishing of the Weyl coefficients $\Psi_{1}=\Psi_{3}=0$ yields the
following conditions on the integration constants: $\beta= \delta$
and $\gamma= - \rho$. By linear transformations of the variable
$p$, $q$, $\tau$ and $\sigma$ the functions $M(p)$ and $N(q)$
reduce to the simple form
\begin{eqnarray}
M(p)=p^2,\,\,\,\,\, \,\,\,\, N(q)=-q^2,
\end{eqnarray}
and then
\begin{eqnarray}
\overline{\Delta}=p^2+q^2=\Delta,
\end{eqnarray}
as before.

Next, we determine the explicit expression of $P(p)$ and $Q(q)$.
Integrating the variable separable equation~(\ref{escalar}) one
gets the following expressions for $P(p)$ and $Q(q)$:
\begin{eqnarray}
\label{P}
P(p)= C_{0}+ C_{1} p -\epsilon p^2 -\Lambda p^4,
\end{eqnarray}
\begin{eqnarray}
\label{Q}
Q(q)= A_{0}+ A_{1} q + \epsilon q^2 - \Lambda q^4.
\end{eqnarray}
Substituting these expressions into equation~(\ref{R01}), one
arrives at the condition $A_{0}=C_{0}$. Finally, entering with
these functions $P(p)$, $Q(q)$, $M(p)$, $N(q)$ and their
derivatives into the equation arising from the vanishing of
$\Psi_{2}$, $\Psi_{2}=0$, one obtains the condition $C_{1}=0$ and
$A_{1}=0$. Therefore, the final expressions for the polynomials
$P(p)$ and $Q(q)$ are
\begin{eqnarray}
\label{PP}
P(p)= C_{0} -\epsilon p^2 -\Lambda p^4,
\end{eqnarray}
\begin{eqnarray}
\label{QQ}
Q(q)= C_{0} + \epsilon q^2 - \Lambda q^4.
\end{eqnarray}
Now we are ready to accomplish the limiting transition from (3+1)
to (2+1) gravity by letting $p \longrightarrow 0$. The metric for
this $p=0$--foliation becomes
\begin{eqnarray}
\label{Plebanski22}
ds_{_{2+1}}^2= \frac{P}{q^2}(d\tau+ q^2 d\sigma)^2+
\frac{q^2}{Q} dq^2  \nonumber \\
- \frac{Q}{q^2} d\tau^2,
\end{eqnarray}
\begin{eqnarray}
\label{Q2}
P=C_{0} , \,\,\,\,\,\,\,\,\, Q(q)= C_{0}+ \epsilon q^2
-\Lambda q^4.
\end{eqnarray}
Again, the integration constant $C_{0}$ can be identified with the
parameter $J^2/4$, $\epsilon$ with $-M$ and $\Lambda$ with
$-1/l^2$. Note also that the resulting $Q(q)/q^2$- function of
expression~(\ref{Q2}) coincides with the squared lapse function
$N(r)$ of equation~(\ref{metrica BTZ}) up to minor redefinitions
($q
\longrightarrow r$, $C_{0} \longrightarrow J^2/4$, $\epsilon 
\longrightarrow -M$,
$\Lambda \longrightarrow -1/l^2$).

We would like to mention that due to symmetry of the
metric~(\ref{Plebanski}), the BTZ metric could be obtained also as
a different real slice of the complexified metric~(\ref{Plebanski})
by  accomplishing there the complex transformation $\tau
\longrightarrow i \tau$ and $\sigma
\longrightarrow i \sigma$ and by setting $m=0$, $l=0$, $e=0$, $\overstar{g}=0$,
together with the slice $q=0$.

We have shown in this work that it is possible to construct spatial
slices of (3+1 ) vacuum plus $\Lambda$ spacetimes, that correspond
to (2+1) vacuum plus $\Lambda$ solutions. In this reduction it is
crucial to require the vanishing of the Weyl tensor components,
i.e. we must demand the (3+1) metric, from which we construct the
spatial slices, to be conformally flat. This is the main
requirement to reduce gravitational fields from (3+1) to ( 2+1)
spacetimes. There are other examples of reductions from vacuum plus
$\Lambda$ term solutions.

The application of a generalization of our criterion to (3+1)
gravitational fields with matter tensors to derive solutions in
(2+1) dimensions with matter sources is under study; in certain
cases the structure of the (3+1) dimensional energy momentum tensor
suffers radical modifications in the transition to (2+1) gravity.
We hope that in a near future we shall be able to report progress
in this lines.

\mbox{} \\
\section*{Acknowledgments}
MC and SdC acknowledge the hospitality of the Physics Department of
CINVESTAV-IPN where this work was done. MC was supported by
COMISION NACIONAL DE CIENCIAS Y TECNOLOGIA through Grant FONDECYT
N$^o$ 1990601, also by Direcci\'{o}n de Promoci\'{o}n y Desarrollo
de la Universidad del B\'{\i}o-B\'{\i}o. SdC was supported from the
COMISION NACIONAL DE CIENCIAS Y TECNOLOGIA through Grant FONDECYT
N$^o$ 1000305 and also from UCV-DGIP 123.744/00. AG was supported
by the Grant N$^o$ 32138E CONACYT-M\'exico.


\begin{thebibliography}{2}
\bibitem{Frolov} V.P. Frolov, S. Hendy and A.L. Larsen, Nucl. Phys. B468, 
336 (1996).

\bibitem{BaTeZa} M. Ba\~nados, C. Teitelboim and J. Zanelli,
Phys. Rev. Lett. 69, 1849 (1992).

\bibitem{PlCa} B. Carter, Commun. Math. Phys. 10, 280 (1968),
idem, Phys. Lett. A 26, 399 (1968); J. Pleba\'nski, Ann. Phys.
(USA) 90, 196 (1975).

\bibitem{GaMa} A. Garc\'\i a and A. Mac\'\i as, Black holes:
Theory and Observation, Lecture notes in Physics (Springer-Verlag,
Berlin, Germany, 1998)

\bibitem{KrStMcHe} D. Kramer, H. Stephani, M. MacCallum and E. Herlt,
Exact Solutions of the Einstein Field Equations. (Deutscher Verlag
der Wissenschaften, Berlin, Germany, 1980).

\end{thebibliography}
\end{document}